\documentclass[11pt,a4paper]{article} 
\pdfoutput=1
\usepackage{dsfont}   
\usepackage{subfigure}
\usepackage{empheq}                                                              
\usepackage{mathtools,amsfonts,amssymb,amsthm}
\usepackage[active]{srcltx}
\usepackage{url}
\usepackage{times,graphicx,xcolor,stackengine}
\usepackage{delarray,fancybox}
\usepackage{mathdots}
\usepackage{eurosym}
\usepackage{relsize}
\usepackage{hyperref}

\usepackage[affil-it]{authblk}

\newcommand{\str}[1]{\overset{\scriptscriptstyle{1/2}}{#1}}
\newcommand{\oti}[1]{\overset{\scriptscriptstyle{1}}{#1}}
\newcommand{\ti}{t_{\iota}}

\newcommand{\tf}{t_{\textit{f}}}

\author{Paolo Muratore-Ginanneschi $^{1}$*, and Kay Schwieger $^{2}$}

\affil{%
$^{1}$ \quad University of Helsinki, Department of Mathematics and Statistics
    P.O. Box 68 FIN-00014, Helsinki, Finland; paolo.muratore-ginanneschi@helsinki.fi\\
$^{2}$ \quad iteratec GmbH, Zettachring 6
70567 Stuttgart, Germany; kay.schwieger@gmail.com}

\title{An application of Pontryagin's principle to Brownian particle engineered equilibration}

\begin{document}
\maketitle
\begin{abstract}
We present a stylized model of controlled equilibration of a small system 
in a fluctuating environment. We derive the equations governing the optimal control 
steering \emph{in finite time} the system between two equilibrium states. The corresponding 
thermodynamic transition is optimal in the sense that occurs at minimum entropy 
if the set of admissible controls is restricted by certain bounds on the time 
derivatives of the protocols. We apply our equations to the engineered equilibration of 
an optical trap considered in a recent proof of principle experiment. We also analyze an elementary model of 
nucleation previously considered by Landauer to discuss the thermodynamic cost of one bit of 
information erasure.
We expect our model to be a useful benchmark for experiment design as it exhibits the same integrability 
properties of well known models of optimal mass transport by a compressible velocity field.
\end{abstract}



\section{Introduction}
\label{sec:intro}

An increasing number of applications in micro and sub-micro scale physics  
call for the development of general techniques for engineered finite-time equilibration 
of systems operating in a thermally fluctuating environment.
Possible concrete examples are the design of nano-thermal engines \cite{BlBe11,RoAbScSiLu14} 
or of micro-mechanical oscillators used for high precision timing or sensing of mass and forces
\cite{LiMeCzShYuSh00}.

A recent experiment \cite{MaPeGuTrCi16} exhibited the feasibility of driving a micro-system 
between two equilibria over a control time several order of magnitude faster than the natural 
equilibration time. The system was a colloidal micro-sphere trapped in 
an optical potential.  There is consensus that non-equilibrium thermodynamics (see e.g. \cite{TrJaRiCrBuLi04})  
of optically trapped micron-sized beads is well captured by Langevin--Smoluchowski 
equations \cite{Jac10}.
In particular, the authors of \cite{MaPeGuTrCi16} took care of showing that it is accurate to conceptualize the 
outcome of their experiment as the evolution of a Gaussian probability density according to a controlled 
Langevin--Smoluchowski dynamics with gradient drift and constant diffusion coefficient. 
Finite time equilibration means that at the end of the control horizon, the probability density 
is solution of the stationary Fokker--Planck equation.
The experimental demonstration consisted in a compression of the confining potential. In such a case,   
the protocol steering the equilibration process is specified by the choice of the time evolution of 
the stiffness of the quadratic potential whose gradient yields the drift in the Langevin--Smoluchowski equation.  
As a result, the set of admissible controls is infinite. The selection  of the control in \cite{MaPeGuTrCi16}
was then based on simplicity of implementation considerations.

A compelling question is whether and how the selection of the protocol may stem from a notion 
of optimal efficiency.  A natural indicator of efficiency in finite-time thermodynamics is
entropy production. 
Transitions occurring at minimum entropy production set a lower bound in Clausius inequality. 
Optimal control of these transitions is, thus, equivalent to a refinement of the second law of 
thermodynamics in the form of an equality.
 
In the Langevin--Smoluchowski framework, entropy production optimal control takes
a particularly simple form if states at the end of the transition are specified by 
sufficiently regular probability densities \cite{AuGaMeMoMG12}. 
Namely, the problem admits an exact mapping into the well
known Monge--Kantorovich optimal mass transport \cite{Villani}. This feature is particularly useful
because the dynamics of the Monge--Kantorovich problem is exactly solvable. Mass transport occurs along 
free-streaming Lagrangian particle trajectories. These trajectories satisfy boundary conditions determined by the map, 
called the Lagrangian map, transforming into each other the data of the problem, 
the initial and the final probability densities.
Rigorous mathematical results \cite{BeBr00,BrFrHeLoMaMoSo03,DePhFi14} preside over the existence, qualitative properties 
and reconstruction algorithms for the Lagrangian map.

The aforementioned results cannot be directly applied to optimal protocols for engineered equilibration. 
Optimal protocols in finite time unavoidably attain minimum entropy by leaving the end probability densities
out of equilibrium. The qualitative reason is that optimization is carried over the
set of drifts sufficiently smooth to mimic all controllable degrees of freedom of the micro-system.
Controllable degrees of freedom are defined as those varying over typical time scales much slower 
than the time scales of Brownian forces \cite{AlRiRi11}.
The set of admissible protocols defined in this way is too large for optimal engineered 
equilibration. The set of admissible controls for equilibration must take into account 
also extra constraints coming from the characteristic time scales of the forces acting on the system. 
From the experimental slant, we expect these restrictions to be strongly contingent on the nature and 
configuration of peripherals in the laboratory setup. From the theoretical point of view, self-consistence 
of Langevin--Smoluchowski modeling imposes a general restriction. The time variation of drift fields controlling the 
dynamics must be slow in comparison to Brownian and inertial forces. 

In the present contribution, we propose a refinement of the entropy production optimal control adapted 
to engineered equilibration. We do this by restricting the set of admissible controls to those
satisfying a non-holonomic constraint on accelerations. The constraint relates the bound on admissible 
accelerations to the pathwise displacement of the system degrees of freedom across the control horizon.  
Such displacement is a deterministic quantity, intrinsically stemming from the boundary conditions inasmuch
we determine it from the Lagrangian map.

This choice of the constraint has several inherent advantages. It yields an intuitive hold on the 
realizability of the optimal process. It also preserves the integrability properties of the unrestricted
control problem specifying the lower bound to the second law. This is so because the bound allows us to
maintain protocols within the admissible set by exerting on them uniform accelerating or decelerating forces.
On the technical side, the optimal control problem can be handled by a direct application of Pontryagin 
maximum principle \cite{Lib12}. For the same reasons as for the refinement 
of the second law \cite{AuGaMeMoMG12}, the resulting optimal control is of deterministic type. This circumstance 
yields a technical simplification but it is not a necessary condition in view of extensions of our approach. We will
return to this point in the conclusions.

The structure of the paper is as follows. In section~\ref{sec:model} we briefly review the Langevin--Smoluchowski 
approach to non-equilibrium thermodynamics \cite{Se98}. This section can be skipped by readers familiar to the topic.
In section~\ref{sec:S} we introduce the problem of optimizing the entropy production. In particular we explain its 
relation with the Schr\"odinger diffusion problem \cite{Sc31,Aebi}. This relation, already pointed out in \cite{PMG13}, 
has recently attracted the attention of mathematicians and probabilists interested in rigorous application of variational principles in
hydrodynamics \cite{ArCrLeZa17}.
In section~\ref{sec:heat-opt} we formulate the Pontryagin principle for our problem. Our main result 
follows in section~\ref{sec:explicit} where we solve in explicit form the optimal protocols. 
Sections~\ref{sec:Gauss} and \ref{sec:Landauer} are devoted to applications. In \ref{sec:Gauss} we revisit 
the theoretical model of the experiment \cite{MaPeGuTrCi16}, the primary motivation of our work.
In section~\ref{sec:Landauer} we apply our results to a stylized model of controlled nucleation obtained 
by manipulating a double well potential. Landauer and Bennett availed themselves of this model 
to discuss the existence of intrinsic thermodynamic cost of computing \cite{Lan61,Ben82}. 
Optimal control of this model has motivated in more recent years 
several theoretical \cite{DiLu09} and experimental works \cite{BeArPeCiDiLu12,KoMaPeAv14,JuGaBe14}. 

Finally, in section~\ref{sec:valley} we compare 
the optimal control we found with those of \cite{AuMeMG12}. This reference applied a regularization technique
coming from instanton calculus \cite{AoKiOkSaWa99} to give a precise meaning to otherwise ill-defined problems in
non-equilibrium thermodynamics, where terminal cost seem to depend on the control rather than being a given function 
of the final state of the system. 

In the conclusions we discuss possible extensions of the present work. The style of the presentation is meant 
to be discursive but relies on notions in between non-equilibrium physics, optimal control theory and probability theory.
For this reason we include in appendices some auxiliary information as a service to the interested reader.

\section{Kinematics and thermodynamics of the model}
\label{sec:model}

We consider a physical process in a $d$-dimensional Euclidean space ($\mathbb{R}^{d}$)
modeled by a Langevin--Smoluchowski dynamics
\begin{eqnarray}
\label{model:LS}
\mathrm{d}\boldsymbol{\xi}_{t}=-\partial_{\boldsymbol{\xi}_{t}}U(\boldsymbol{\xi}_{t},t)\,\mathrm{d}t
+\sqrt{\frac{2}{\beta}}\,\mathrm{d}\boldsymbol{\omega}_{t}
\end{eqnarray}
The stochastic differential $\mathrm{d}\boldsymbol{\omega}_{t}$  stands here for
the increment of a standard $d$-dimensional Wiener process at time $t$ \cite{Jac10}. 
$U\colon\mathbb{R}^{d}\otimes\mathbb{R}\mapsto\mathbb{R}$ denotes a smooth scalar potential
and $\beta^{-1}$ is a constant sharing the same canonical dimensions as $U$. 
We also suppose that the initial state of the system is specified by a smooth 
probability density
\begin{eqnarray}
\label{model:init}
\mathrm{P}(\boldsymbol{q}\leq \boldsymbol{\xi}_{\ti}<\boldsymbol{q}+\mathrm{d}\boldsymbol{q})
=\mathrm{p}_{\iota}(\boldsymbol{q})\mathrm{d}^{d}\boldsymbol{q}
\end{eqnarray}
Under rather general hypotheses, the Langevin--Smoluchowski equation 
(\ref{model:LS}) can be derived as the scaling limit of
the overdamped non-equilibrium dynamics of a classical system weakly coupled to an 
heat bath \cite{Zwa01}. The Wiener process in (\ref{model:LS}) thus embodies 
thermal fluctuations of order $\beta^{-1}$.
The fundamental simplification entailed by (\ref{model:LS}) 
is the possibility to establish a framework of elementary relations linking the dynamical 
to the statistical levels of description of a non-equilibrium process \cite{Se98,LeSp99}. 
In fact, the kinematics of (\ref{model:LS}) ensures that for any time-autonomous, confining 
potential the dynamics tends to a unique Boltzmann equilibrium state.
\begin{eqnarray}
\mathrm{p}_{\mathrm{eq}}(q)\propto\exp\Big{(}-\beta\,U(\boldsymbol{q})\Big{)} 
\nonumber
\end{eqnarray}
Building on the foregoing observations \cite{Se98}, we may then identify $U$ over a finite time horizon 
with the internal energy of the system. The differential of $U$ 
\begin{eqnarray}
\mathrm{d}U(\boldsymbol{\xi}_{t},t)=\mathrm{d}t\,\partial_{t}U(\boldsymbol{\xi}_{t},t)
+\mathrm{d}\boldsymbol{\xi}_{t}\str{\cdot}\partial_{\boldsymbol{\xi}_{t}}U(\boldsymbol{\xi}_{t},t)
\label{model:diff}
\end{eqnarray} 
yields the energy balance in the presence of thermal fluctuations due to interactions with the
environment. We use the notation $\str{\cdot}$ for the Stratonovich differential \cite{Jac10}.
From (\ref{model:diff}) we recover the first law of thermodynamics by averaging over 
the realizations of the Wiener process. In particular, we interpret 
\begin{eqnarray}
\label{od:work}
\mathcal{W}=\mathrm{E}\int_{t_{o}}^{t_{f}}\mathrm{d}t\,\partial_{t}U(\boldsymbol{\xi}_{t},t)
\end{eqnarray}
as the average work done on the system. Correspondingly, 
\begin{eqnarray}
\label{od:heat}
\mathcal{Q}=-\mathrm{E}\int_{t_{o}}^{t_{f}}\mathrm{d}\boldsymbol{\xi}_{t}\str{\cdot}
\partial_{\boldsymbol{\xi}_{t}}U(\boldsymbol{\xi}_{t},t)
\end{eqnarray}
is the average heat discarded by the system into the heat bath and therefore
\begin{eqnarray}
\mathcal{W}-\mathcal{Q}
=\operatorname{E}\Big{(}U(\boldsymbol{\xi}_{\tf},\tf)-U(\boldsymbol{\xi}_{\tf},\tf)\Big{)}
\label{model:1stlaw}
\end{eqnarray}
is the embodiment of the first law.

The kinematics of stochastic processes \cite{Nelson01}, allow us also to write 
a meaningful expression for the second law of thermodynamics.
The expectation value of a Stratonovich differential is in general amenable to the form
\begin{eqnarray}
\label{model:heat}
\mathcal{Q}=-\mathrm{E}\int_{\ti}^{t_{f}}\mathrm{d}t\,(\boldsymbol{v}\cdot
\partial_{\boldsymbol{\xi}_{t}}U)(\boldsymbol{\xi}_{t},t)
\end{eqnarray}
where
\begin{eqnarray}
\label{model:cf}
\boldsymbol{v}(\boldsymbol{q},t)
=-\partial_{\boldsymbol{q}}\left(U(\boldsymbol{q},t)+\frac{1}{\beta}\ln \mathrm{p}(\boldsymbol{q},t)\right)
\end{eqnarray}
is the current velocity. For a potential drift, the current velocity vanishes 
identically at equilibrium. As well known from stochastic mechanics \cite{Fen52,Nelson85}, 
the current velocity permits to couch the Fokker--Planck equation into the form of a 
deterministic mass transport equation. Hence, upon observing that
\begin{eqnarray}
\label{}
\mathrm{E}\int_{\ti}^{t_{f}}\mathrm{d}t\,(\boldsymbol{v}\cdot\partial_{\boldsymbol{\xi}_{t}}\ln \mathrm{p})
(\boldsymbol{\xi}_{\tf},\tf)=
\mathrm{E}\int_{\ti}^{t_{f}}\mathrm{d}t\left(\partial_{t}+
\boldsymbol{v}_{t}\cdot\partial_{\boldsymbol{\xi}_{t}}\right)\ln \mathrm{p}(\boldsymbol{\xi}_{\tf},\tf)
=\mathrm{E}\ln\frac{\mathrm{p}(\boldsymbol{\xi}_{\tf},\tf)}{\mathrm{p}(\boldsymbol{\xi}_{\ti},\ti)}
\end{eqnarray}
we can recast (\ref{model:heat}) into the form 
\begin{eqnarray}
\label{model:2ndlaw}
\mathcal{Q}_{T}=\mathcal{Q}-\frac{1}{\beta}
\mathrm{E}\ln\frac{\mathrm{p}(\boldsymbol{\xi}_{\tf},\tf)}{\mathrm{p}(\boldsymbol{\xi}_{\ti},\ti)}
=\mathrm{E}\int_{\ti}^{t_{f}}\mathrm{d}t\, \parallel\boldsymbol{v}(\boldsymbol{\xi}_{t},t)\parallel^{2}
\end{eqnarray}
which we interpret as the second law of thermodynamics (see e.g. \cite{Qia01}). 
Namely, if we define $\mathcal{E}=\beta\,\mathcal{Q}_{T}$ as the total entropy change in 
$[\ti\,\tf]$, (\ref{model:2ndlaw}) states that the sum of the entropy generated by heat 
released into the environment plus the change of the Gibbs--Shannon entropy of the system 
is positive definite and vanishes only at equilibrium. 
The second law in the form  (\ref{model:2ndlaw}) immediately implies a bound on the average 
work done on the system. To evince this fact, we avail us of the equality
\begin{eqnarray}
\label{od:heat-cv}
\mathcal{W}=\mathrm{E}\left(U(\boldsymbol{\xi}_{\tf},\tf)-U(\boldsymbol{\xi}_{\ti},\ti)
+\frac{1}{\beta}\ln\frac{\mathrm{p}(\boldsymbol{\xi}_{\tf},\tf)}{\mathrm{p}(\boldsymbol{\xi}_{\ti},\ti)}\right)
+\mathcal{Q}_{T}
\end{eqnarray}
and define the current velocity potential
\begin{eqnarray}
F(\boldsymbol{q},t)=U(\boldsymbol{q},t)+\frac{1}{\beta}\ln \mathrm{p}(\boldsymbol{q},t)
\nonumber
\end{eqnarray}
We then obtain
\begin{eqnarray}
\mathcal{W}=\mathrm{E}\Big{(}F(\boldsymbol{\xi}_{\tf},\tf)-F(\boldsymbol{\xi}_{\ti},\ti)\Big{)}
+\mathcal{Q}_{T}\geq \mathrm{E}\Big{(}F(\boldsymbol{\xi}_{\tf},\tf)-F(\boldsymbol{\xi}_{\ti},\ti)\Big{)}
\nonumber
\end{eqnarray}
In equilibrium thermodynamics the Helmholtz free energy is defined as the difference
\begin{eqnarray}
\mathcal{F}=\mathcal{U}-\beta^{-1}\,\mathcal{S}
\nonumber
\end{eqnarray}
between the internal energy $\mathcal{U}$ and entropy $\mathcal{S}$ of a system at temperature $\beta^{-1}$. 
This relation admits a non-equilibrium extension 
by noticing that the information content \cite{Sha48} of the system probability density
\begin{eqnarray}
S(\boldsymbol{q},t)=-\ln \mathrm{p}(\boldsymbol{q},t)
\nonumber
\end{eqnarray}
weighs the contribution of individual realizations of (\ref{model:LS}) to the Gibbs-Shannon entropy. 
We refer to \cite{Nelson01} for the kinematic and thermodynamic interpretation of the information
content as osmotic potential. We also emphasize that the notions above can be given an intrinsic meaning
using the framework of stochastic differential geometry \cite{Nelson85,PMG13}.
Finally, it is worth noticing that the above relations can be regarded as a 
special case of macroscopic fluctuation theory \cite{BeDeGaJoLa15}.

\section{Non-equilibrium thermodynamics and Schr\"odinger diffusion}
\label{sec:S}

We are interested in thermodynamic transitions between an initial state (\ref{model:init}) at time $\ti$
and a \emph{pre-assigned final state} at time also specified by a smooth probability density
\begin{eqnarray}
\label{S:final}
\mathrm{P}(\boldsymbol{q}\leq \boldsymbol{\xi}_{\tf}<\boldsymbol{q}+\mathrm{d}\boldsymbol{q})
=\mathrm{p}_{\mathit{f}}(\boldsymbol{q})\mathrm{d}^{d}\boldsymbol{q}
\end{eqnarray}
We also suppose that the cumulative distribution functions of (\ref{model:init}) and (\ref{S:final})
are related by a Lagrangian map $\boldsymbol{\ell}\colon\mathbb{R}^{d}\mapsto\mathbb{R}^{d}$ such that
\begin{eqnarray}
\mathrm{P}(\boldsymbol{\xi}_{\ti}<\boldsymbol{q})=\mathrm{P}(\boldsymbol{\xi}_{\ti}<\boldsymbol{\ell}(\boldsymbol{q}))
\label{S:Lagrangian}
\end{eqnarray}
According to the Langevin--Smoluchowski dynamics (\ref{model:LS}), the evolution of probability densities
obey a Fokker--Planck equation, a first order in time partial differential equation. 
As a consequence, a price we pay to steer transitions 
between assigned states is to regard the drift in (\ref{model:LS}) not as an assigned quantity 
but as a control. 
A priori a control is only implicitly characterized by the set of conditions which make it admissible. 
Informally speaking, admissible controls are all those drifts steering the process 
$\left\{\boldsymbol{\xi}_{t},t\in[\ti\,,\tf]\right\}$ between the assigned end states (\ref{model:init}) and 
(\ref{S:final}) while ensuring that at any time $t\in [\ti\,,\tf]$ the Langevin--Smoluchowski dynamics 
remains well-defined.

Schr\"odinger \cite{Sc31} considered already in 1931 the problem of controlling a diffusion process between
assigned states. His work was motivated by the quest of a statistical interpretation of quantum mechanics.
In modern language \cite{Da91,RoTh02}, the problem can be rephrased as follows. Given 
(\ref{model:init}) and (\ref{S:final}) and a reference diffusion process, determine the diffusion process
interpolating between (\ref{model:init}) and (\ref{S:final}) while minimizing the value of its
Kullback--Leibler divergence (relative entropy) \cite{KuLe51} with respect to the reference process.
A standard application (appendix~\ref{ap:KL}) of Girsanov formula \cite{Jac10} shows that the Kullback--Leibler 
divergence of (\ref{model:LS}) with respect to the Wiener process is
\begin{eqnarray}
\mathcal{K}(\mathrm{P} \parallel\mathrm{P}_{\boldsymbol{\omega}})
=\frac{\beta}{2}\operatorname{E}\int_{\ti}^{\tf}\mathrm{d}t\,\|\partial_{\boldsymbol{\xi}_{t}}U(\boldsymbol{\xi}_{t},t)\|^{2}
\label{S:KL}
\end{eqnarray}
$\mathrm{P}$ and $\mathrm{P}_{\boldsymbol{\omega}}$ denote respectively the measures of the
process solution of (\ref{model:LS}) with drift $-\partial_{\boldsymbol{q}}U(\boldsymbol{q},t)$ and of the
Wiener process $\boldsymbol{w}$. The expectation value on the right hand side is with respect to $\mathrm{P}$
as elsewhere in the text. 
A now well-established result in optimal control theory see e.g. \cite{Da91,RoTh02} is that the optimal
value of the drift satisfies a backward Burgers equation with terminal condition specified by the
solution of the Beurling--Jamison integral equations. We refer to \cite{Da91,RoTh02} for further details.
What interest us here is to emphasize the \emph{analogy} with the problem of minimizing the entropy production
$\mathcal{E}$ in a transition between assigned states.

Several observations are in order at this stage.

The first observation is that also (\ref{model:2ndlaw}) can be directly interpreted as a Kullback--Leibler
divergence between two probability measures. Namely, we can write (appendix~\ref{ap:KL})
\begin{eqnarray}
\mathcal{K}(\operatorname{P}\parallel \operatorname{P}_{R})=
\frac{\beta}{2}\operatorname{E}\int_{\ti}^{\tf}\mathrm{d}t\,\|\boldsymbol{v}(\boldsymbol{\xi}_{t},t)\|^{2}
\label{S:ep}
\end{eqnarray}
for $\operatorname{P}_{R}$ the path-space measure of the process
\begin{eqnarray}
\label{S:backward}
\mathrm{d}\boldsymbol{\xi}_{t}=\partial_{\boldsymbol{\xi}_{t}}U(\boldsymbol{\xi}_{t},t)\,\mathrm{d}t
+\sqrt{\frac{2}{\beta}}\,\mathrm{d}\boldsymbol{\omega}_{t}
\end{eqnarray}
evolving backward in time from the final condition (\ref{S:final}) \cite{JiangQianQian,ChGa08}.

The second observation has more far reaching consequences for optimal control. 
The entropy production depends upon the drift of (\ref{model:LS}) exclusively through the 
current velocity (\ref{model:cf}). Hence we can treat the current velocity itself as natural 
control quantity for (\ref{S:ep}). This fact entails major simplifications \cite{AuGaMeMoMG12}.
The current velocity can be thought as deterministic rather than stochastic velocity field (see 
\cite{Nelson01} and appendix~\ref{ap:cv}).
Thus, we can couch the optimal control of (\ref{S:ep}) into the problem of minimizing 
the kinetic energy of a classical particle traveling from an initial position
$\boldsymbol{q}$ at time $\ti$ and a final position $\boldsymbol{\ell}(\boldsymbol{q})$ 
at time $\tf$ specified by the Lagrangian map $\boldsymbol{\ell}$ (\ref{S:Lagrangian}). In other words, entropy 
production minimization in the Langevin--Smoluchowski framework is equivalent to solve a 
classical optimal transport problem \cite{Villani}.

The third observation comes as a consequence of the second one. The optimal value of the entropy production 
is equal to the Wasserstein distance \cite{JoKiOt98} between the initial and final 
probability measures of the system, see \cite{Gaw13} for details. This fact yields a simple 
characterization of the Landauer bound and permits a fully explicit analysis of the 
thermodynamics of stylized isochoric micro-engines (see \cite{PMGSc14} and refs therein). 

Finally, the construction of Schr\"odinger diffusions via optimal control of (\ref{S:KL}) corresponds 
to a \emph{viscous regularization} of the optimal control equations occasioned by the Schr\"odinger diffusion
problem (\ref{S:ep}).

\section{Pontryagin's principle for bounded accelerations}
\label{sec:heat-opt}

An important qualitative feature of the solution of the optimal control of the entropy 
production is that the system starts from (\ref{model:init}) and reaches (\ref{S:final}) 
with non-vanishing current velocity. This means that the entropy production attains a 
minimum value when the end-states of the transition are out-of-equilibrium.  
We refer to this lower bound as the refinement of the second law.

Engineered equilibration transitions are, however, subject to at least
two  further  types of  constraints  not  taken  into account  in  the
derivation of the refined second law.  The first type of constraint is
on the  set of admissible  controls. For example,  admissible controls
cannot vary  in an  arbitrary manner:  the fastest  time scale  in the
Langevin--Smoluchowski  dynamics is  set  by the  Wiener process.  The
second type  is that  end-states are  at equilibrium.  In mathematical
terms, this means that the current velocity must vanish identically at
$\ti$ and $\tf$.

We formalize a deterministic control problem modeling these constraints.
Our goal is to minimize the functional
\begin{eqnarray}
\mathcal{E}=\int_{\ti}^{t_{f}}\mathrm{d}t\,\beta\,\|\boldsymbol{\nu}_{t}\|^{2}
\label{heat-opt:cost}
\end{eqnarray}
over the set of trajectories generated for any given choice of the measurable 
control $\boldsymbol{\alpha}_{t}$ by the differential equation
\begin{subequations}
\label{heat-opt:eqs}
\begin{align}
\label{heat-opt:eqs1}
\boldsymbol{\dot{\chi}}_{t} &= \boldsymbol{\nu}_{t}
\\
\label{heat-opt:eqs2}
\boldsymbol{\dot{\nu}}_{t} &= \boldsymbol{\alpha}_{t}
\end{align}
\end{subequations} 
satisfying the boundary conditions
\begin{eqnarray}
\label{heat-opt:bc1}
\boldsymbol{\chi}_{\ti}=\boldsymbol{q}
\hspace{1.0cm}\&\hspace{1.0cm}
\boldsymbol{\chi}_{\tf}=\boldsymbol{\ell}(\boldsymbol{q})
\end{eqnarray}
We dub the dynamical variable $\chi_{t}$ \emph{running Lagrangian map} as it describes the 
evolution of the Lagrangian map within the control horizon. 
We restrict the set of admissible controls $\mathbb{A}=\left\{\alpha_{t},t\in[\ti\,,\tf]\right\}$ 
to those enforcing equilibration at the boundaries of the control 
horizon
\begin{eqnarray}
\label{heat-opt:bc2}
\boldsymbol{\nu}_{\ti}=0
\hspace{1.0cm}\&\hspace{1.0cm}
\boldsymbol{\nu}_{\tf}=0
\end{eqnarray}
whilst satisfying the bound
\begin{eqnarray}
|\alpha_{t}^{(i)}|\leq \frac{K^{(i)}(\boldsymbol{q})}{(\tf-\ti)^{2}}
\hspace{1.0cm}\forall\,t\in[\ti\,,\tf]\hspace{1.0cm}\forall\,i=1,\dots,d
\label{heat-opt:bound}
\end{eqnarray}
We suppose that the $K^{(i)}(\boldsymbol{q})\,>\,0$ $i=1,\dots,d$ are strictly positive functions 
of the \emph{initial data} $\boldsymbol{q}$ of the form
\begin{eqnarray}
\label{heat-opt:bound1}
K^{(i)}(\boldsymbol{q})\propto |\ell^{(i)}(\boldsymbol{q})-q^{(i)}|
\end{eqnarray}
The constraint is non-holonomic inasmuch it depends on the 
initial data of a trajectory. The proportionality (\ref{heat-opt:bound1}) relates
the bound on acceleration to the Lagrangian displacement needed to satisfy 
the control problem.

We resort to Pontryagin principle \cite{Lib12} to find normal extremals 
of (\ref{heat-opt:cost}).  We defer the statement of Pontryagin principle as well as the 
discussion of abnormal extremals to appendix~\ref{ap:Pontryagin}.
We proceed in two steps. We first avail us of Lagrange multipliers to define the 
effective cost functional
\begin{eqnarray}
\mathcal{A}=\int_{\ti}^{t_{f}}\mathrm{d}t\,\Big{(}\beta\,\parallel\boldsymbol{\nu}_{t}\parallel^{2}
+\boldsymbol{\eta}_{t}\cdot\left(\boldsymbol{\dot{\chi}}_{t}-\boldsymbol{\nu}_{t}\right)
+\boldsymbol{\theta}_{t}\cdot\left(\boldsymbol{\dot{\nu}}_{t}-\boldsymbol{\alpha}_{t}\right)
\Big{)}
\nonumber
\end{eqnarray}
subject to the boundary conditions (\ref{heat-opt:bc1}), (\ref{heat-opt:bc2}). 
Then, we couch the cost functional into an explicit Hamiltonian form 
\begin{eqnarray}
\mathcal{A}=\int_{\ti}^{t_{f}}\mathrm{d}t\,\Big{(}
\boldsymbol{\eta}_{t}\cdot\boldsymbol{\dot{\chi}}_{t}
+\boldsymbol{\theta}_{t}\cdot\boldsymbol{\dot{\nu}}_{t}
-H(\boldsymbol{\chi}_{t},\boldsymbol{\nu}_{t},\boldsymbol{\eta}_{t},\boldsymbol{\theta}_{t},\boldsymbol{\alpha}_{t})
\Big{)}
\label{heat-opt:heat}
\end{eqnarray}
with
\begin{eqnarray}
H(\boldsymbol{\chi}_{t},\boldsymbol{\nu}_{t},\boldsymbol{\eta}_{t},\boldsymbol{\theta}_{t},\boldsymbol{\alpha}_{t})=
\boldsymbol{\eta}_{t}\cdot\boldsymbol{\nu}_{t}
+\boldsymbol{\theta}_{t}\cdot\boldsymbol{\alpha}_{t}
-\beta\,\parallel\boldsymbol{\nu}_{t}\parallel^{2}
\nonumber
\end{eqnarray}
Pontryagin's principle yields a rigorous proof of the intuition that extremals of 
the optimal control equations correspond to stationary curves of the action (\ref{heat-opt:heat}) 
with Hamiltonian
\begin{eqnarray}
H_{\star}(\boldsymbol{\chi}_{t},\boldsymbol{\nu}_{t},\boldsymbol{\eta}_{t},\boldsymbol{\theta}_{t})=
\max_{\boldsymbol{\alpha}\in\mathbb{A}}H(\boldsymbol{\chi}_{t},\boldsymbol{\nu}_{t},\boldsymbol{\eta}_{t},
\boldsymbol{\theta}_{t},\boldsymbol{\alpha}_{t})=
\boldsymbol{\eta}_{t}\cdot\boldsymbol{\nu}_{t}
+\frac{\sum_{i=1}^{d}K^{(i)}|\theta_{t}^{(i)}|}{(\tf-\ti)^{2}}
-\beta\,\parallel\boldsymbol{\nu}_{t}\parallel^{2}
\nonumber
\end{eqnarray}
In view of the boundary conditions (\ref{heat-opt:bc1}), (\ref{heat-opt:bc2}) extremals satisfy 
the Hamilton system of equations formed by (\ref{heat-opt:eqs1}) and
\begin{subequations}
\label{heat-opt:opteqs}
\begin{eqnarray}
\label{heat-opt:opteqs1}
\dot{\nu}^{(i)}_{t}=\partial_{\boldsymbol{\theta}_{t}}H_{\star}=\frac{K^{(i)}}{(\tf-\ti)^{2}}\operatorname{sgn}\theta_{t}^{(i)}
\end{eqnarray}
\begin{eqnarray}
\label{heat-opt:opteqs2}
\dot{\boldsymbol{\eta}}_{t}=-\partial_{\boldsymbol{\chi}_{t}}H_{\star}=0
\end{eqnarray}
\begin{eqnarray}
\label{heat-opt:opteqs3}
\dot{\boldsymbol{\theta}}_{t}=-\partial_{\boldsymbol{\nu}_{t}}H_{\star}=-\boldsymbol{\eta}_{t}+2\,\beta\,\boldsymbol{\nu}_{t}
\end{eqnarray}
\end{subequations}
In writing (\ref{heat-opt:opteqs1}) we adopt the convention
\begin{eqnarray}
\operatorname{sgn}0=0
\nonumber
\end{eqnarray}

\section{Explicit solution in the $1d$ case}
\label{sec:explicit}

The extremal equations (\ref{heat-opt:eqs1}), (\ref{heat-opt:opteqs}) are time-autonomous and 
do not couple distinct vector components. It is therefore non-restrictive to focus on the $d=1$
case in the time horizon $[0,T]$.

The Hamilton equations are compatible with two behaviors. 
A ``push-region'' where the running Lagrangian map variable evolves 
with constant acceleration
\begin{eqnarray}
\ddot{\chi}_{t}=\frac{K}{T^{2}}\operatorname{sgn}\theta_{t}\hspace{1.0cm}\&\hspace{1.0cm}\theta_{t}\neq 0
\nonumber
\end{eqnarray}
and a ``no-action'' region specified by the conditions
\begin{eqnarray}
\theta_{t}=0 \hspace{1.0cm}\&\hspace{1.0cm}-\boldsymbol{\eta}_{\star}+2\,\beta\,\boldsymbol{\nu}_{\star}=0
\label{explicit:na}
\end{eqnarray}
where $\chi_{t}$ follows a free streaming trajectory:
\begin{eqnarray}
\dot{\chi}_{t}=\boldsymbol{\nu}_{\star}
\nonumber
\end{eqnarray}
We call \emph{switching times} the values of $t$ corresponding to the boundary values 
of a no-action region. Switching times correspond to discontinuities of the acceleration $\alpha_{t}$.
Drawing from the intuition offered by the solution of the unbounded acceleration case,
we compose push and no-action regions to construct a single solution trajectory satisfying 
the boundary conditions. If we surmise that during the control horizon only two switching times
occur, we obtain
\begin{eqnarray}
\label{explicit:vel}
\nu_{t}=
\left\{
\begin{array}{ll}
\dfrac{K}{T^{2}}\, t \, \operatorname{sgn}\theta_{0}
\hspace{0.5cm}&\hspace{0.5cm} 
t\,\in\,[0,t_{1})
\\[0.3cm]
\dfrac{K\,t_{1} }{T^{2}} \, \operatorname{sgn}\theta_{0}
\hspace{0.5cm}&\hspace{0.5cm} 
t\,\in\,[t_{1},t_{2}]
\\[0.3cm]
\dfrac{K }{T^{2}}
\,\left(t_{1}\,\operatorname{sgn}\theta_{0}+(t-t_{2})\,\operatorname{sgn}\theta_{T}\right)
\hspace{0.5cm}&\hspace{0.5cm} 
t\,\in\,(t_{2},T]
\end{array}
\right.
\end{eqnarray}
which implies
\begin{eqnarray}
\theta_{t}=
\left\{
\begin{array}{ll}
\theta_{0}-\dfrac{\beta\,K \, t\,(2\,t_{1}-t)}{T^{2}}\, \operatorname{sgn}\theta_{0}
\hspace{0.5cm}&\hspace{0.5cm} 
t\,\in\,[0,t_{1})
\\[0.3cm]
0
\hspace{0.5cm}&\hspace{0.5cm} 
t\,\in\,[t_{1},t_{2}]
\\[0.3cm]
\dfrac{K\,(t-t_{2})^{2} }{T^{2}}\,\operatorname{sgn}\theta_{T}
\hspace{0.5cm}&\hspace{0.5cm} 
t\,\in\,(t_{2},T]
\end{array}
\right.
\label{explicit:theta}
\end{eqnarray}
Self-consistence of the solutions fixes the initial data in (\ref{explicit:theta})
\begin{eqnarray}
\theta_{0}=\dfrac{\beta\,K \, t_{1}^{2}}{T^{2}}\, \operatorname{sgn}\theta_{0}
\nonumber
\end{eqnarray}
whilst the requirement of vanishing velocity at $t=T$ determines the relation between the
switching times
\begin{eqnarray}
t_{2}=T+\frac{\operatorname{sgn}\theta_{0}}{\operatorname{sgn}\theta_{T}}t_{1}
\nonumber
\end{eqnarray} 
Self-consistence then dictates
\begin{eqnarray}
\operatorname{sgn}\theta_{\tf}=-\operatorname{sgn}\theta_{t_{0}}
\nonumber
\end{eqnarray}
We are now ready to glean the information we unraveled by solving (\ref{heat-opt:opteqs}),
to write the solution of (\ref{heat-opt:eqs1})
\begin{eqnarray}
\label{explicit:sol1}
\chi_{t}=q+
\left\{
\begin{array}{ll}
\dfrac{K \, t^{2}}{2\,T^{2}}\, \operatorname{sgn}\theta_{0}
\hspace{0.5cm}&\hspace{0.5cm} 
t\,\in\,[0,t_{1})
\\[0.3cm]
\dfrac{K\,t_{1} \,(2\,t-t_{1})}{2\,T^{2}} \, \operatorname{sgn}\theta_{0}
\hspace{0.5cm}&\hspace{0.5cm} 
t\,\in\,[t_{1},T-t_{1}]
\\[0.3cm]
K \dfrac{2\,t_{1}\,(T-t_{1})-(T-t)^{2}}{2\,T^{2}}\operatorname{sgn}\theta_{0}
\hspace{0.5cm}&\hspace{0.5cm} 
t\,\in\,(T-t_{1},T]
\end{array}
\right.
\end{eqnarray}
The terminal condition on $\chi_{t}$ fixes the values of 
$t_{1}$ and $\operatorname{sgn}\theta_{t_{0}}$:
\begin{eqnarray}
\ell(q)=q+ \dfrac{K(q)\,t_{1}\,(T-t_{1})}{T^{2}}\operatorname{sgn}\theta_{t_{0}}
\nonumber
\end{eqnarray}
The equation for $t_{1}$ well posed only if
\begin{eqnarray}
\operatorname{sgn}\theta_{t_{0}}=\operatorname{sgn}\Big{(}\ell(q)-q\Big{)}
\label{explicit:sol2}
\end{eqnarray}
The only admissible solution is then of the form
\begin{eqnarray}
t_{1}=\frac{T}{2}\left(1-\sqrt{ 1-4\,\delta}\right)
\label{explicit:sol3}
\end{eqnarray}
The switching time is independent of $q$ in view of (\ref{heat-opt:bound1}). It is realizable 
as long as
\begin{eqnarray}
\delta=\frac{|\ell(q)-q|}{K(q)}\,\leq\,\frac{1}{4}\hspace{1.0cm}\forall\,q\,\in\,\mathbb{R}
\label{explicit:realizability}
\end{eqnarray}
The threshold value of $\delta$ correspond to the acceleration needed to construct an optimal protocol 
consisting of two push regions matched at half control horizon.

\subsection{Qualitative properties of the solution}
\label{sec:qualitative}

Equation (\ref{explicit:sol1}) complemented by (\ref{explicit:sol2}) and 
the realizability bound (\ref{explicit:realizability})
fully specify the solution of the optimization problem we set out to solve.
The solution is optimal because it is obtained by composing locally optimal 
solutions.
Qualitatively, it states that transitions between equilibrium states are possible
at the price of the formation of symmetric boundary layers determined by the occurrence 
of the switching times. For $\delta\ll 1$ the relative size of the boundary layers is
\begin{eqnarray}
\frac{t_{1}}{T}=\frac{T-t_{2}}{T}\approx \delta
\nonumber
\end{eqnarray}
In the same limit, the behavior of the current velocity far from the boundaries 
tends to the optimal value of the refined second law \cite{AuGaMeMoMG12}. 
Namely, for $t\in [t_{1}\,,t_{f}]$ we find
\begin{eqnarray}
\dfrac{K(q)\,t_{1} }{T^{2}} \,\operatorname{sgn}\Big{(}\ell(q)-q\Big{)}\overset{\delta \ll 1}{\approx}
\dfrac{K(q)\,\delta }{T}\operatorname{sgn}\Big{(}\ell(q)-q\Big{)}=\frac{\ell(q)-q}{T}
\nonumber
\end{eqnarray}
More generally for any $0\leq \,t_{1}\,\leq T/2$, we can couch (\ref{explicit:sol1}) into the form
\begin{eqnarray}
\label{qualitative:sol}
\chi_{t}=q+\Big{(}\ell(q)-q\Big{)}\times
\left\{
\begin{array}{ll}
\dfrac{t^{2} }{2\,t_{1}\,(T-t_{1})}
\hspace{0.5cm}&\hspace{0.5cm} 
t\,\in\,[0,t_{1})
\\[0.3cm]
\dfrac{2\,t-t_{1}}{2\,(T-t_{1})} \,
\hspace{0.5cm}&\hspace{0.5cm} 
t\,\in\,[t_{1},T-t_{1}]
\\[0.3cm]
\left(1-\dfrac{(T-t)^{2}}{2\,t_{1}\,(T-t_{1})}\right)
\hspace{0.5cm}&\hspace{0.5cm} 
t\,\in\,(T-t_{1},T]
\end{array}
\right.
\end{eqnarray}
The use of the value of the switching time $t_{1}$ to parametrize the bound
simplifies the derivation of the Eulerian representation of the current velocity. 
Namely, in order to find the field $v\colon\mathbb{R}\times[0,T]\mapsto\mathbb{R}$ satisfying
\begin{eqnarray}
\nu_{t}=v(\chi_{t},t)
\label{qualitative:Euler}
\end{eqnarray}
we can invert (\ref{qualitative:sol}) by taking advantage of the fact that all the arguments
of the curly brackets are independent of the position variable $q$.

We also envisage that the representation (\ref{qualitative:sol}) may be of use to analyze 
experimental data when finite measurement resolution may affect the precision with 
which microscopic forces acting on the system are known.

\section{Comparison with experimental swift engineering (ESE) protocols}
\label{sec:Gauss}

The experiment reported in \cite{MaPeGuTrCi16} showed that a micro-sphere immersed
in water and trapped in an optical harmonic potential can be driven in finite time from 
an equilibrium state to another. The probability distribution of the particle in and out 
equilibrium remained Gaussian within experimental accuracy.

It is therefore expedient to describe more in detail the solution of the optimal
control problem in the case when the initial equilibrium distribution 
in one dimension is normal, i.e. Gaussian with zero mean and variance $\beta^{-1}$. We also 
assume that the final equilibrium state is Gaussian and satisfy (\ref{S:Lagrangian}) with
Lagrangian map
\begin{eqnarray}
\ell(q)=\sigma\,q+h
\nonumber
\end{eqnarray}
The parameters $h$ and $\sigma$ respectively describe a change of the mean and of 
the variance of the distribution. 
We apply (\ref{S:Lagrangian}) and (\ref{qualitative:sol}) for any $t\in [0,T]$
to derive the minimum entropy production evolution of the probability density.
In consequence of (\ref{heat-opt:bound1}), the running Lagrangian map leaves Gaussian 
distributions invariant in form with mean value
\begin{eqnarray}
\label{}
\operatorname{E}\xi_{t}=h\times
\left\{
\begin{array}{ll}
\dfrac{t^{2} }{2\,t_{1}\,(T-t_{1})}
\hspace{0.5cm}&\hspace{0.5cm} 
t\,\in\,[0,t_{1})
\\[0.3cm]
\dfrac{(2\,t-t_{1})}{2\,(T-t_{1})} \,
\hspace{0.5cm}&\hspace{0.5cm} 
t\,\in\,[t_{1},T-t_{1}]
\\[0.3cm]
\dfrac{2\,t_{1}\,(T-t_{1})-(T-t)^{2}}{2\,t_{1}\,(T-t_{1})}
\hspace{0.5cm}&\hspace{0.5cm} 
t\,\in\,(T-t_{1},T]
\end{array}
\right.
\end{eqnarray}
and variance
\begin{eqnarray}
\label{}
\operatorname{V}\xi_{t}=
\begin{cases}
 \dfrac{\left(2 \,t_{1}\, (T-t_{1})+(\sigma -1)\, t^{2}\right)^2}{4 \,\beta\,  t_{1}^2\, (T-t_{1})^2} & t\in [0,t_{1})
\\[0.3cm]
 \dfrac{\Big{(}2\,(T-t_{1})+(\sigma -1) (2\,t-t_{1})\Big{)}^2}{4 \,\beta\,(T-t_{1})^2} & t\in [t_{1},T-t_{1}] 
\\[0.3cm]
 \dfrac{\Big{(}2\, t_{1}\,(T-t_{1})+(\sigma -1) (2\,t_{1}\,(T-t_{1})-(T-t)^{2})\Big{)}^2}{4 \,\beta \, t_{1}^{2}\,
   \,(T-t_{1})^2} & t\in (T-t_{1},T]
\end{cases}
\end{eqnarray}
Finally, we find that the Eulerian representation (\ref{qualitative:Euler}) of the current 
velocity at $\chi_{t}=q$ is 
\begin{eqnarray}
\label{}
v(q,t)
=
\begin{cases}
 \dfrac{2\, t\, (h+q\, (\sigma -1))}{2 \,t_{1} \,(T-t_{1})+(\sigma -1)\, t^2} & t\in [0,t_{1})
\\[0.3cm]
 \dfrac{2\,\Big{(}h+q (\sigma -1)\Big{)}}{2\,(T-t_{1})+(\sigma -1)\, (2\,t-t_{1})} & t\in [t_{1},T-t_{1}] 
\\[0.3cm]
 \dfrac{2 (T-t) \Big{(}h+q (\sigma -1)\Big{)}}{2 \,t_{1}\,(T-t_{1})+(\sigma -1)(2\,t_{1}\,(T-t_{1})- (T-t)^{2})} &
  t\in (T-t_{1},T]
\end{cases}
\end{eqnarray}
The foregoing expression allows us to write explicit expressions for the all the thermodynamic quantities
governing the energetics of the optimal transition. In particular, the minimum entropy production is
\begin{eqnarray}
\label{}
\operatorname{E}\int_{0}^{\tf}\mathrm{d}t\,v^{2}(\xi_{t},t)=\frac{T\,(3\,T-4\,t_{1})}{3\,(T-t_{1})^{2}}\mathcal{E}_{\infty}
\end{eqnarray}
with
\begin{eqnarray}
\label{}
\mathcal{E}_{\infty}=\frac{h^{2}\,\beta+(\sigma-1)^{2}}{\beta\,T}
\end{eqnarray}
the value of the minimum entropy production appearing in the refinement of the second law \cite{AuGaMeMoMG12}. 
\begin{figure}[ht]
\centering
\subfigure[Work (continuous curve, blue on-line) and heat release 
(dashed curve, yellow on-line) 
during the control horizon. Inset: time evolution of the  
variance of the process]{
\stackinset{r}{4pt}{l}{8pt}{\includegraphics[width=2.0cm]{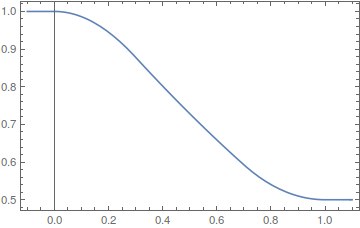}}{\includegraphics[width=5.0cm]{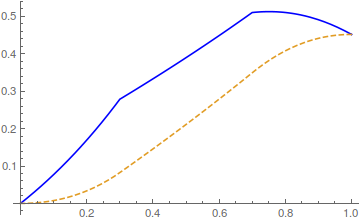}}
\label{fig:11}
}
\hspace{1.0cm}
\subfigure[Entropy production (continuous curve, blue on-line) and heat release 
(dashed curve, yellow on-line) during the control horizon]{
\includegraphics[width=5.0cm]{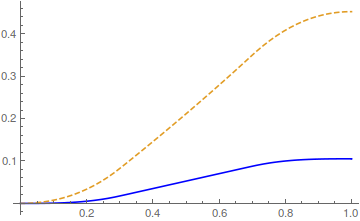}
  \label{fig:12}
}
\caption{First fig.~\ref{fig:11} and second law fig.~\ref{fig:12} of thermodynamics for the 
same transition between Gaussian states 
as in \cite{MaPeGuTrCi16}.
The initial state is a normal distribution with variance $\beta^{-1}$. The final distribution is 
Gaussian with variance $\beta^{-1}/2$.
The condition $K(q)\propto |\ell(q)-q|$ ensures that the probability 
density remains Gaussian at any time in the
control horizon. The proportionality factor is chosen such that $t_{1}=0.3$ in (\ref{qualitative:sol}).
The behavior of the variance (inset of fig~\ref{fig:11}) is qualitatively identical to the 
one observed in \cite{MaPeGuTrCi16} Fig.~2.
The behavior of the average work and heat also reproduces the one of Fig.~3 of \cite{MaPeGuTrCi16}.
}
\label{fig:1}
\end{figure}
In Fig.~\ref{fig:1} we plot the evolution of the running average values of the work done on the system, the heat release
and the entropy production during the control horizon. In particular, Fig.~\ref{fig:11} illustrates 
the first law of thermodynamics during the control horizon. 
A transition between Gaussian equilibrium states occurs without any change in the internal energy of the 
system. The average heat and work must therefore coincide at the end of the control horizon. 
The theoretical results are consistent with the experimental results of 
\cite{MaPeGuTrCi16}.

\section{Optimal controlled nucleation and Landauer bound}
\label{sec:Landauer}

The form of the bound (\ref{heat-opt:bound1}) and running Lagrangian map formula 
(\ref{qualitative:sol}) reduce the computational cost of the solution the optimal entropy 
production control to the determination of the Lagrangian map (\ref{S:Lagrangian}). 
In general, the conditions presiding to the qualitative properties of the Lagrangian 
map have been studied in depth in the context of optimal mass transport \cite{Villani}. 
We refer to \cite{DePhFi14} and \cite{Gaw13} respectively for a self-contained overview 
from respectively the mathematics an physics slant.

For illustrative purpose, we revisit here the stylized 
model of nucleation analyzed in \cite{AuGaMeMoMG12}. 
Specifically, we consider the transition between two equilibria in one dimension. 
The initial state is described by the symmetric double well:
\begin{eqnarray}
\mathrm{p}_{\iota}(q)=Z_{\iota}^{-1}\exp-\beta\,\frac{(q^{2}-\bar{q}^{2})^{2}}{\sigma^{2}}
\nonumber
\end{eqnarray}
In the final state the probability is concentrated around a single minimum of the potential:
\begin{eqnarray}
\mathrm{p}_{\mathit{f}}(q)=Z_{\mathit{f}}^{-1}
\exp-\beta\frac{(q-\bar{q})^{2}\Big{(}(q-\bar{q})+\bar{q}\,(3\,q-\bar{q})\Big{)}}{\sigma^{2}}
\nonumber
\end{eqnarray}
In the foregoing expressions $\sigma$ is a constant ensuring consistency of 
the canonical dimensions. 

\begin{figure}
\centering
\subfigure[Boundary conditions for the nucleation problem]{
\includegraphics[width=4.0cm]{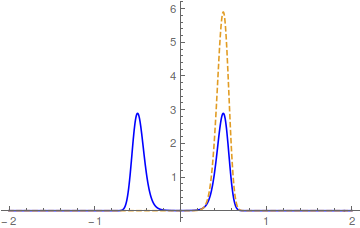}
 \label{fig:21}
}
\hspace{0.3cm}
\subfigure[Lagrangian map]{
\includegraphics[width=4.0cm]{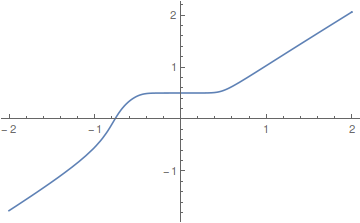}
 \label{fig:22}
}
\hspace{0.3cm}
\subfigure[Numerical derivative of the Lagrangian map]{
\includegraphics[width=4.0cm]{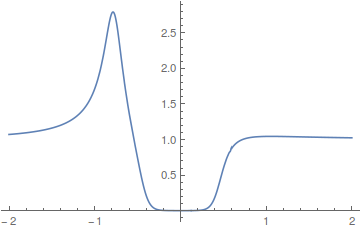} 
\label{fig:23}
}
\caption{
Initial (solid curve, blue on-line) and final  (dashed curve, blue on-line) 
probability distribution of the state of the system for $\beta=112$ $\sigma=1$
and $\bar{q}=1/2$. The evaluation of the Lagrangian map occasions numerical stiffness 
in the region in between the two minima. 
}
\label{fig:2}
\end{figure}
We used the ensuing elementary algorithm to numerically determine the 
Lagrangian map.  We first computed the  
median $z(1)$ of the assigned probability distributions and then evaluated first the left and then right 
branch   of the Lagrangian map. For the left branch, we proceeded iteratively in $z(k)$ as follows
\begin{itemize}
\item[\textbf{Step 1}] We renormalized the distribution restricted to $[-\infty, z(k)]$.
\item[\textbf{Step 2}] We computed the $0.9$ quantile $z(k+1) < z(k)$ of the remaining distribution.
\item[\textbf{Step 3}] We solved the ODE 
\begin{eqnarray}
\frac{\mathrm{d}\ell }{\mathrm{d} q}=\frac{\mathrm{p}_{\iota}(q)}{\mathrm{p}_{\mathit{f}}(\ell(q))}
\nonumber
\end{eqnarray}
\end{itemize}
We skipped \textbf{Step 3} whenever the difference $|z(k) - z(k-1)|$ turned out to be smaller than 
a given threshold 'resolution'. We plot the results of this computation in Fig.~2.

\begin{figure}
\begin{minipage}[b]{1.0\linewidth}
\centering
\subfigure[$t=0.05$]{
\includegraphics[width=4.0cm]{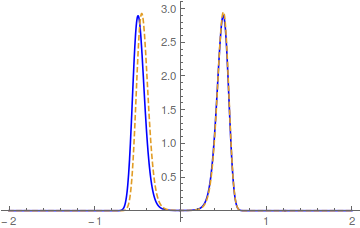}
}
\hspace{0.2cm}
\subfigure[$t=0.25$]{
\includegraphics[width=4.0cm]{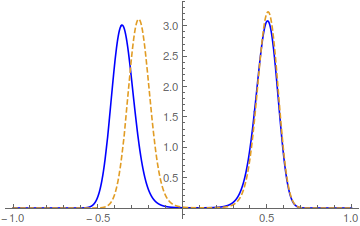}
}
\hspace{0.2cm}
\subfigure[$t=0.45$]{
\includegraphics[width=4.0cm]{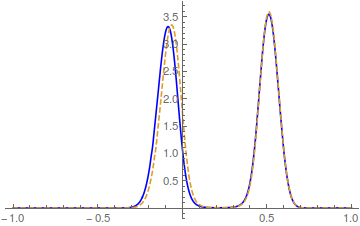}
}
\end{minipage}
\begin{minipage}[b]{1.0\linewidth}
\centering
\subfigure[$t=0.65$]{
\includegraphics[width=4.0cm]{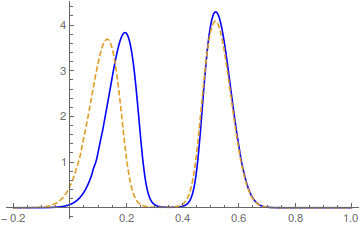}
}
\hspace{0.2cm}
\subfigure[$t=0.85$]{
\includegraphics[width=4.0cm]{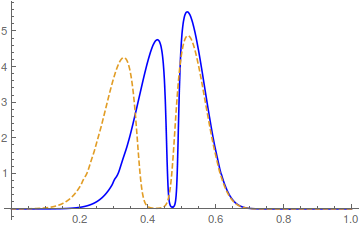}
}
\hspace{0.2cm}
\subfigure[$t=0.98$]{
\includegraphics[width=4.0cm]{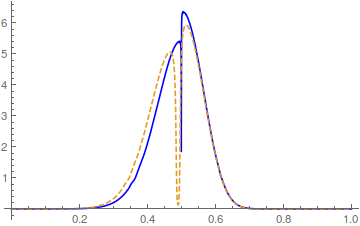}
}
\end{minipage}%
\caption{Probability density snapshots at different times within the control horizon. The plots are for $T=1$ and 
switching time $t_{1}=10^{-6}$ (dashed interpolation curve, yellow on-line) and $t_{1}=0.3$ 
(continuous interpolation curve, blue on-line) $\bar{q}=0.5$, $\sigma=1$ and $\beta=112$. We plot 
the Lagrangian map in the interval $q\in [-2\,,2]$}
\label{fig:3}
\end{figure}

\begin{figure}
\begin{minipage}[b]{1.0\linewidth}
\centering
\subfigure[$t=0.05$]{
\includegraphics[width=4.0cm]{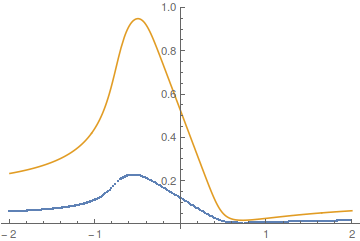}
}
\hspace{0.2cm}
\subfigure[$t=0.25$]{
\includegraphics[width=4.0cm]{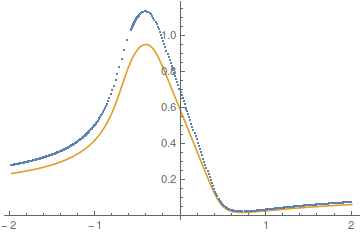}
}
\hspace{0.2cm}
\subfigure[$t=0.45$]{
\includegraphics[width=4.0cm]{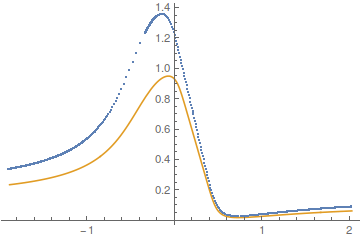}
}
\end{minipage}
\begin{minipage}[b]{1.0\linewidth}
\centering
\subfigure[$t=0.65$]{
\includegraphics[width=4.0cm]{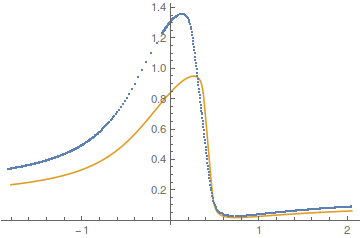}
}
\hspace{0.2cm}
\subfigure[$t=0.85$]{
\includegraphics[width=4.0cm]{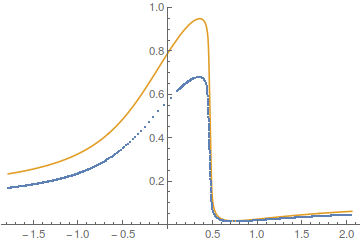}
}
\hspace{0.2cm}
\subfigure[$t=0.98$]{
\includegraphics[width=4.0cm]{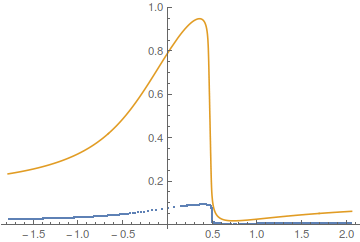}
}
\end{minipage}%
\caption{Current velocity snapshots at different times within the control horizon. The plots are for $T=1$ and 
switching time $t_{1}=10^{-6}$ (continuous interpolation, yellow on-line) and $t_{1}=0.3$ (points, blue on-line)}
\label{fig:4}
\end{figure}

Once we know the Lagrangian map, we can numerically evaluate the running Lagrangian 
map (\ref{qualitative:sol}) and its spatial derivatives. In Fig.~\ref{fig:3} we report the evolution 
of the probability density in the control horizon for two reference values of the switching 
time.
Fig.~\ref{fig:4} illustrates the the corresponding evolution of the current velocity.
The qualitative behavior is intuitive. The current velocity starts and ends with vanishing value,
it catches up with the value for $t_{1}\downarrow 0$, i.e. when the bound on acceleration tends to infinity, 
in the bulk of the control horizon. There the displacement described by the running Lagrangian 
map occurs at speed higher than in the $t_{1}\downarrow 0$ case. The overall value of the entropy production is 
always higher than in the $t_{1}\downarrow 0$ limit.
From (\ref{qualitative:sol}) we can also write the running values 
of average heat released by the system.  The running average heat is
\begin{eqnarray}
Q(t)=
-\frac{1}{\beta}\int_{\mathbb{R}}\mathrm{d}^{d}q\,\mathrm{p}_{\iota}(q)\,\ln \frac{\mathrm{d} \chi_{t}(q)}{\mathrm{d} q}
+ \int_{0}^{t}\mathrm{d}s\,\int_{\mathbb{R}}\mathrm{d}^{d}q\,\mathrm{p}_{\iota}(q)\,\nu_{t}^{2}(q)
\nonumber
\end{eqnarray}
and the running average work
\begin{eqnarray}
W(t)=\int_{\mathbb{R}}\mathrm{d}q\,\mathrm{p}_{\iota}(q)\,
F(\chi_{t}(q),t)
+ \int_{0}^{t}\mathrm{d}s\,\int_{\mathbb{R}}\mathrm{d}^{d}q\,\mathrm{p}_{\iota}(q)\,\nu_{t}^{2}(q)
\nonumber
\end{eqnarray}
with
\begin{eqnarray}
F(\chi_{t}(q),t)=-\int_{0}^{q}\mathrm{d}y\frac{\mathrm{d} \chi_{t}}{\mathrm{d} y}(y)\,\nu_{t}(y)
-\frac{1}{\beta}\int_{\mathbb{R}}\mathrm{d}^{d}q\,\mathrm{p}_{\iota}(q)\,\ln \frac{\mathrm{d} \chi_{t}(q)}{\mathrm{d} q}
\label{Landauer:free_en}
\end{eqnarray}
The second summand on the right hand side of (\ref{Landauer:free_en}) fixes the arbitrary constant in the 
Helmholtz potential in the same way as in the Gaussian case. 
\begin{figure}
\centering
\hspace{0.3cm}
\subfigure[First law of thermodynamics for the optimally controlled nucleation. Continuous curve (blue on-line) 
running average work. Dashed curve (yellow on-line) running average heat]{
\includegraphics[width=5.0cm]{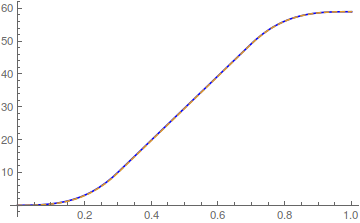}
 \label{fig:51}
}
\hspace{0.5cm}
\subfigure[ Running average entropy production. The continuous curve (blue on-line) 
is obtained for switching time at $t_{1}=0.3$, the dashed curve (yellow on-line) for $t_{1}=10^{-6}$]{
\includegraphics[width=5.0cm]{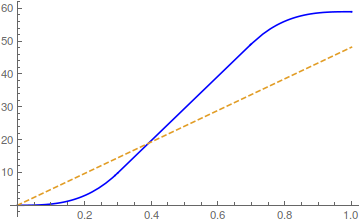} 
\label{fig:52}
}
\caption{
First and second law of thermodynamics for the optimally controlled nucleation transition. 
All parameters are as in Fig.~\ref{fig:2}. 
The qualitative picture is the same as in the Gaussian case Fig.~\ref{fig:1} with the running average 
work above the running average heat. The numerical values yield, however, almost overlapping curves.  
The running average entropy production in fig.~\ref{fig:52} is strictly monotonic in the control horizon. 
The entropy production rate vanishes at the boundary highlighting the reach of an equilibrium state 
when the switching time is $t_{1}=0.3$.
}
\label{fig:5}
\end{figure}
In Fig.~\ref{fig:5} we plot the running average work, heat and entropy production.

\section{Comparison with the valley method regularization}
\label{sec:valley}

An alternative formalism to study transitions between equilibrium states in the Langevin--Smoluchowski limit 
was previously proposed in \cite{AuMeMG12}. As in the present case, \cite{AuMeMG12} takes advantage of the 
possibility to map the stochastic optimal control problem into a deterministic one via the current velocity
formalism. Physical constraints on admissible controls are, however, enforced by adding to the
entropy production rate a penalty term proportional to the squared current acceleration. In terms of the
entropy production functional (\ref{heat-opt:cost}) we can couch the regularized functional of \cite{AuMeMG12}
into the form 
\begin{eqnarray}
\mathcal{A}=\mathcal{E}+\varepsilon\,\tau^{2}\, \|\delta_{\chi} \mathcal{E}\|^{2}
\nonumber
\end{eqnarray}
$\delta_{\chi}\mathcal{E}$ stands for the variation of $\mathcal{E}$ with respect to the running Lagrangian map.
The idea behind the approach is the ``valley method''  
advocated by \cite{AoKiOkSaWa99} for instanton calculus. The upshot is to approximate field
configurations satisfying boundary conditions incompatible with stationary values of classical variational principles 
by adding extra terms to the action functional. The extra term is proportional to the squared first variation of the 
classical action. Hence it vanishes whenever there exists 
a classical field configurations matching the desired boundary conditions. It otherwise raises the order of the time 
derivative in the problem thus permitting to satisfy extra boundary conditions. 

Optimal control problems are well-posed if terminal costs are pure functionals of the boundary conditions. 
The rationale for considering valley method regularized thermodynamic functionals is to give a non-ambiguous 
meaning to the optimization of functionals whenever naive formulations of 
the problem yield boundary conditions or terminal costs as functional of the controls.
  
\begin{figure}
\centering
\hspace{0.3cm}
\subfigure[$\frac{\chi_{t}-q}{\ell(q)-q}$]{
\includegraphics[width=5.0cm]{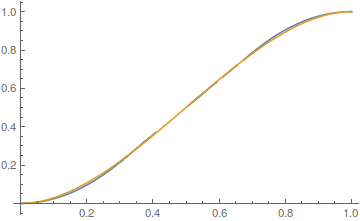}
 \label{fig:61}
}
\hspace{0.5cm}
\subfigure[Running entropy production]{
\includegraphics[width=5.0cm]{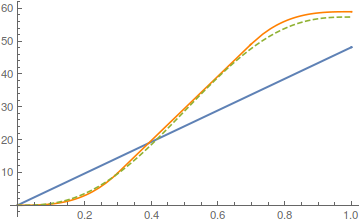} 
\label{fig:62}
}
\caption{Qualitative comparison of universal part of the running Lagrangian maps 
(\ref{qualitative:sol}) (continuous curve, blue on line) and (\ref{valley:rlm}) (dashed curve, orange on line) 
fig.~\ref{fig:61}.
In (\ref{valley:rlm}) we choose $\tau=1$, $\varepsilon=0.3$. Fig.~\ref{fig:62} evinces, as to be expected, 
the qualitatively equivalent behaviors of the entropy production for finite value ($t_{1}=0.3$) 
of the switching time. The dashed green line is computed from (\ref{valley:rlm}). The continuous blue line
is the lower bound for the transition as predicted by \cite{AuGaMeMoMG12}.
}
\label{fig:6}
\end{figure}

Contrasted with the approach proposed in the present work, \cite{AuMeMG12} has one evident drawback and one edge. 
The drawback is that the quantities actually minimized are not anymore the original thermodynamic functionals.
The edge is that the resulting optimal protocol has better analyticity properties.
In particular, the running Lagrangian map takes the form 
\begin{eqnarray}
\chi_{t}=q+\frac{\ell(q)-q}{T-2\,\tau\,\sqrt{ \varepsilon}\,\tanh\frac{T}{2\,\tau\,\sqrt{ \varepsilon}}}
\left(t-\tau\,\sqrt{\varepsilon}\,
\frac{\sinh\frac{2\,t-T}{2\,\tau\,\sqrt{ \varepsilon}}+\sinh\frac{T}{2\,\tau\,\sqrt{ \varepsilon}}}
{\cosh\frac{T}{2\,\tau\,\sqrt{ \varepsilon}}}
\right)
\label{valley:rlm}
\end{eqnarray}
In fig.~\ref{fig:61} we compare the qualitative behavior of the universal part of the running Lagrangian 
map predicted by the valley method and by the bound (\ref{heat-opt:bound}) on admissible current accelerations. 
The corresponding values of the running average entropy production are in fig.~\ref{fig:62}.

The upshot of the comparison is the weak sensitivity of the optimal protocol to the detail
of the optimization once the intensity of the constraint on the admissible control (i.e. the current acceleration)
is fixed. We believe that this is an important observation for experimental applications 
(see, e.g., discussion in the conclusions of 
\cite{JuGaBe14}) as the detail of how control parameters can be turned on and off in general depend on the detailed 
laboratory setup and on the restrictions by the available peripherals.

\section{Conclusions and outlooks}
\label{sec:conclusion}

We presented a stylized model of engineered equilibration of a micro-system. Owing to explicitly integrability
modulo numerical reconstruction of the Lagrangian map, we believe that our model may provide an useful benchmark
for the devising of efficient experimental setups. Furthermore extensions of the
current model are possible although at the price of some complications.

The first extension concerns the form of the constraint imposed on admissible protocols. Here we showed that
choosing the current acceleration constraint in the form (\ref{heat-opt:bound1}) greatly simplifies the determination
of the switching times. It also guarantees that optimal control with only two switching times exists for all boundary
conditions if we allow accelerations to take sufficiently large values. The non-holonomic form of the
constraint  (\ref{heat-opt:bound}) may turn out to be restrictive for the study 
of transitions for which admissible controls are specified by given forces. If the current velocity
formalism is still applicable to these cases, then the design of optimal control still follows the steps we described 
here. In particular, uniformly accelerated Lagrangian displacement at the end of the control horizon correspond to 
the first terms of the integration of Newton law in Peano--Picard series. The local form of the acceleration 
may then occasion some qualitative differences in the form of the running Lagrangian map. 
Furthermore, the analysis of the realizability conditions of the optimal control may also become 
more involved.

A further extension is optimal control when constraints on admissible controls are imposed
directly on the drift field appearing in the stochastic evolution equation. Constraints of this type are
natural when inertial effects become important and the dynamics is governed by Langevin--Kramers equation 
in the so-called under-damped approximation. In the Langevin--Kramers framework, finding minimum entropy production
thermodynamic transitions requires instead a full-fledged formalism of stochastic optimal control \cite{PMGSc14}. 
Nevertheless, it is possible also in that case to proceed in a way analogous to one of the present paper 
by applying the stochastic version of Pontryagin principle \cite{Bi78,KoPa93,Rog01}. 

We expect that considering these theoretical refinements will be of interest in view of the increasing 
available experimental resolution for efficient design of atomic force microscopes 
\cite{LiMeCzShYuSh00,CuMaPeGuTrCi16}.

\section{Acknowledgments}

The authors thank S.~Ciliberto for useful discussions.
The work of KS was mostly performed during his stay at the department of Mathematics and Statistics 
of the University of Helsinki. 
PMG acknowledges support from Academy of Finland via the Centre of Excellence in Analysis 
and Dynamics Research (project No. 271983) and to the AtMath collaboration at the University of Helsinki. 

\appendix
\section*{Appendices}

\section{Evaluation of Kullback--Leibler divergences}
\label{ap:KL}

Let us consider first the drift-less process
\begin{eqnarray}
\label{ap:Wiener}
\mathrm{d}\boldsymbol{\xi}_{t}=
\sqrt{\frac{2}{\beta}}\,\mathrm{d}\boldsymbol{\omega}_{t}
\end{eqnarray}
with initial data (\ref{model:init}). If we denote by $\mathrm{P}_{\omega}$ the 
path-space Wiener measure generated by (\ref{ap:Wiener}) in $[\ti\,,\tf]$, 
Girsanov formula yields
\begin{eqnarray}
\frac{\mathrm{d}\mathrm{P}}{\mathrm{d}\mathrm{P}_{\omega}}=
\exp-\frac{\beta}{2}\int_{\ti}^{\tf}\left(\mathrm{d}\boldsymbol{\xi}_{t}\cdot\partial_{\boldsymbol{\xi}_{t}}U
+\mathrm{d}t\dfrac{\|\partial_{\boldsymbol{\xi}_{t}}U\|^{2}}{2}\right)
\nonumber
\end{eqnarray}
The Kullback--Leibler divergence is defined as
\begin{eqnarray}
\mathcal{K}(\mathrm{P}||\mathrm{P}_{\omega})
=\operatorname{E}\int_{\ti}^{\tf}\ln\frac{\mathrm{d}\mathrm{P}}{\mathrm{d}\mathrm{P}_{\omega}}
\nonumber
\end{eqnarray}
The expectation value is with respect the measure $\mathrm{P}$
generated by (\ref{model:LS}):
\begin{eqnarray}
\lefteqn{
\mathcal{K}(\mathrm{P}\parallel\mathrm{P}_{\omega})
=-\frac{\beta}{2}\operatorname{E}\int_{\ti}^{\tf}\left(\mathrm{d}\boldsymbol{\xi}_{t}\cdot\partial_{\boldsymbol{\xi}_{t}}U
+\mathrm{d}t\dfrac{\|\partial_{\boldsymbol{\xi}_{t}}U\|^{2}}{2}\right)
}
\nonumber\\&&
=-\frac{\beta}{2}\operatorname{E}\int_{\ti}^{\tf}\left(
(\mathrm{d}\boldsymbol{\xi}_{t}+\mathrm{d}t\partial_{\boldsymbol{\xi}_{t}}U)\cdot\partial_{\boldsymbol{\xi}_{t}}U
-\mathrm{d}t\dfrac{\|\partial_{\boldsymbol{\xi}_{t}}U\|^{2}}{2}\right)
\nonumber
\end{eqnarray} 
The last expression readily recovers (\ref{S:KL}) as $\mathrm{d}\boldsymbol{\xi}_{t}
+\mathrm{d}t\,\partial_{\boldsymbol{\xi}_{t}}U$
is a Wiener process with respect to $\mathrm{P}$.

To show that the entropy production is proportional to the 
Kullback--Leibler divergence between the path-space measures of
(\ref{model:LS}) and (\ref{S:backward}) we observe that
\begin{eqnarray}
\frac{\mathrm{d}\mathrm{P}_{R}}{\mathrm{d}\mathrm{P}_{\omega}}=
\exp\frac{\beta}{2}\int_{\ti}^{\tf}\left(\mathrm{d}\boldsymbol{\xi}_{t}\oti{\cdot}\partial_{\boldsymbol{\xi}_{t}}U
-\mathrm{d}t\dfrac{\|\partial_{\boldsymbol{\xi}_{t}}U\|^{2}}{2}\right)
\label{ap:RN}
\end{eqnarray}
The stochastic integral is evaluated in the post-point prescription
as the Radon--Nikodym derivative between backward processes must be a martingale 
with respect the filtration of future event (see e.g. \cite{Mey82} for an 
elementary discussion). We then avail us of the time reversal invariance of the Wiener 
process to write
\begin{eqnarray}
\lefteqn{
\frac{\mathrm{d}\mathrm{P}}{\mathrm{d}\mathrm{P}_{R}}=\frac{\mathrm{p}_{\iota}(\boldsymbol{\xi}_{\ti})}{
\mathrm{p}_{\mathit{f}}(\boldsymbol{\xi}_{\tf})}
\exp-\beta\int_{\ti}^{\tf}\left(\mathrm{d}\boldsymbol{\xi}_{t}\str{\cdot}\partial_{\boldsymbol{\xi}_{t}}U
\right)
}
\nonumber\\&&
=
\exp-\beta\int_{\ti}^{\tf}\left(\mathrm{d}\boldsymbol{\xi}_{t}
\str{\cdot}\partial_{\boldsymbol{\xi}_{t}}\left(U+\frac{1}{\beta}\ln \mathrm{p}\right)
+\mathrm{d}t\,\partial_{t}\ln \mathrm{p}\right)
\nonumber
\end{eqnarray}
Finally, the definition
\begin{eqnarray}
\mathcal{K}(\mathrm{P}\parallel \mathrm{P}_{R})=
\operatorname{E}\int_{\ti}^{\tf}\ln\frac{\mathrm{d}\mathrm{P}}{\mathrm{d}\mathrm{P}_{R}}
\nonumber
\end{eqnarray}
recovers (\ref{S:ep}) since probability conservation entails
\begin{eqnarray}
\operatorname{E}\partial_{t}\ln \mathrm{p}=0
\nonumber
\end{eqnarray}
whilst the properties of the Stratonovich integral \cite{Nelson85} yield
\begin{eqnarray}
\operatorname{E}\int_{\ti}^{\tf}\mathrm{d}\boldsymbol{\xi}_{t}
\str{\cdot}\partial_{\boldsymbol{\xi}_{t}}\left(U+\frac{1}{\beta}\ln \mathrm{p}\right)
=-\operatorname{E}\int_{\ti}^{\tf}\mathrm{d}t\,\|\boldsymbol{v}\|^{2}
\nonumber
\end{eqnarray}
We refer to e.g. \cite{LeSp99,MaReMo00,JiangQianQian,ChGa08} for thorough discussions
of the significance and applications of the entropy production in stochastic models of 
non-equilibrium statistical mechanics and to \cite{GrPuSa12,GrPuSaViVu13} for applications to 
non-equilibrium fluctuating hydrodynamics and granular materials.

\section{Current velocity and acceleration in terms of the generator of the stochastic process }
\label{ap:cv}

The current velocity is the conditional expectation along the realizations of (\ref{model:LS}) 
of the time symmetric conditional increment
\begin{eqnarray}
\boldsymbol{v}(\boldsymbol{q},t)=
\lim_{\tau\downarrow 0}\frac{\operatorname{E}\Big{(}\boldsymbol{\xi}_{t+\tau}
-\boldsymbol{\xi}_{t-\tau}\Big{|}\boldsymbol{\xi}_{t}=\boldsymbol{q}\Big{)}}{2\,\tau}
\nonumber
\end{eqnarray} 
A relevant feature of the time symmetry is that the differential can be regarded as the result of
the action of a generator including only first order derivatives in space:
\begin{eqnarray}
\boldsymbol{v}(\boldsymbol{\xi}_{t},t)=\bar{\mathds{D}}_{\boldsymbol{\xi}_{t}}\boldsymbol{\xi}_{t}
\nonumber
\end{eqnarray} 
where
\begin{eqnarray}
\bar{\mathds{D}}_{\boldsymbol{\xi}_{t}}:=\frac{\mathds{D}_{\boldsymbol{\xi}_{t}}+\mathds{D}_{\boldsymbol{\xi}_{t}}^{*}}{2}
\label{cv:symmetric}
\end{eqnarray}
On the right hand side of (\ref{cv:symmetric}) there appear the scalar generator of (\ref{model:LS})
\begin{eqnarray}
\mathds{D}_{\boldsymbol{q}}=\partial_{t}-(\partial_{\boldsymbol{q}}U)(\boldsymbol{q},t)\cdot\partial_{\boldsymbol{q}}+\frac{1}{\beta}\partial_{\boldsymbol{q}}^{2}
\nonumber
\end{eqnarray}
and the generator of the dual process conjugated by time-reversal of the probability 
density in $[\ti\,,\tf]$ \cite{Nelson01,Nelson85}:
\begin{eqnarray}
\mathds{D}_{\boldsymbol{q}}^{*}=\partial_{t}
-(\partial_{\boldsymbol{q}}U+\frac{2}{\beta}\partial_{\boldsymbol{q}}\ln \mathrm{p})(\boldsymbol{q},t)\cdot\partial_{\boldsymbol{q}}
-\frac{1}{\beta}\partial_{\boldsymbol{q}}^{2}
\nonumber
\end{eqnarray}
The arithmetic averages of these generators readily defines a first order differential 
operator as in the deterministic case. Analogously, we define the current acceleration as
\begin{eqnarray}
\boldsymbol{a}(\boldsymbol{q},t)=\lim_{\tau\downarrow 0}\frac{\operatorname{E}\Big{(}\boldsymbol{v}(\boldsymbol{\xi}_{t+\tau},t+\tau)
-\boldsymbol{v}(\boldsymbol{\xi}_{t-\tau},t-\tau)\Big{|}\boldsymbol{\xi}_{t}=\boldsymbol{q}\Big{)}}{2\,\tau}
\nonumber
\end{eqnarray} 
or equivalently
\begin{eqnarray}
\boldsymbol{\alpha}_{t}=\boldsymbol{a}(\boldsymbol{\xi}_{t},t)=\bar{\mathds{D}}_{\boldsymbol{\xi}_{t}}^{2}\boldsymbol{\xi}_{t}
\nonumber
\end{eqnarray}

\section{Pontryagin principle}
\label{ap:Pontryagin}

We recall the statement of Pontryagin's principle for fixed time and
fixed boundary conditions \cite{AgSa04,Lib12}.

\noindent\textbf{Maximum Principle}:\emph{
Let the functional 
\begin{eqnarray}
\mathcal{A}=\int_{\ti}^{\tf}\mathrm{d}t\,L(\boldsymbol{\xi}_{t},\boldsymbol{\alpha}_{t},t)
\label{Pontryagin:action}
\end{eqnarray}
be subject to the dynamical constraint 
\begin{eqnarray}
\boldsymbol{\dot{\xi}}_{t}=\boldsymbol{b}(\boldsymbol{\xi}_{t},\boldsymbol{\alpha}_{t},t)
\label{Pontryagin:ode}
\end{eqnarray}
and the endpoint constraints 
\begin{eqnarray}
\boldsymbol{\xi}_{\ti}=\boldsymbol{q}_{\iota}
\hspace{1.0cm}\&\hspace{1.0cm}
\boldsymbol{\xi}_{\tf}=\boldsymbol{q}_{\mathit{f}}
\nonumber
\end{eqnarray}
with the parameter $\alpha_{t}$ belonging for fixed $t$ to a set $\mathrm{U}\subseteq \mathbb{R}^{n}$, 
the variable $\boldsymbol{\xi}_{t}$ taking values in $\mathbb{R}^{d}$ or in a open subset $\mathrm{X}$ of $\mathbb{R}^{d}$ 
and the time interval $[\ti\,,\tf]$ fixed. 
A necessary condition for a function $\boldsymbol{\bar{\alpha}}_{t}\colon[\ti\,,\tf]\mapsto\mathrm{U}$ and a corresponding
solution $\boldsymbol{\bar{\xi}}_{t}$ of (\ref{Pontryagin:ode}) to solve the minimization of (\ref{Pontryagin:action})
is that there exist a function t $\boldsymbol{\bar{\pi}}_{t}\colon[\ti\,,\tf]\mapsto\mathbb{R}^{d}$  
and a constant $p_{o}\leq 0$ such that
\begin{itemize}
\item $(\boldsymbol{\bar{\pi}}_{t},\bar{p}_{0})\neq (\boldsymbol{0},0)$ $\forall\,t\in[\ti\,,\tf]$ (non-triviality condition)
\item for each fixed $t$
\begin{eqnarray}
H_{\star}(\boldsymbol{q},\boldsymbol{p},p_{0}t)=\max_{\boldsymbol{a}\in \mathrm{U}}
\Big{(}\boldsymbol{p}\cdot\boldsymbol{b}(\boldsymbol{q},\boldsymbol{a},t)+p_{0}\,L(\boldsymbol{q},\boldsymbol{a},t)\Big{)}
\nonumber
\end{eqnarray}
(maximum condition)
\item $(\boldsymbol{\bar{\xi}}_{t},\boldsymbol{\bar{\pi}}_{t})$ obey the equations
\begin{eqnarray}
\bar{\dot{\bar{\xi}}}_{t}=\partial_{\boldsymbol{\bar{\pi}}_{t}}H_{\star}(\boldsymbol{\bar{\xi}}_{t},\boldsymbol{\bar{\pi}}_{t}.\bar{p}_{0},t)
\hspace{1.0cm}\&\hspace{1.0cm}
\bar{\dot{\bar{\pi}}}_{t}=-\partial_{\boldsymbol{\bar{\xi}}_{t}}H_{\star}(\boldsymbol{\bar{\xi}}_{t},\boldsymbol{\bar{\pi}}_{t},\bar{p}_{0},t)
\nonumber
\end{eqnarray}
(Hamilton system condition)
\end{itemize}
}
The proof of the maximum principles requires subtle topological considerations culminating with the application
of Brouwer's fixed point theorem. The maximum principle has, nevertheless, an intuitive content.
Namely, we can reformulate the problem in an extended configuration space by adding the ancillary equation
\begin{subequations}
\label{}
\begin{eqnarray}
\label{}
\dot{\zeta}_{t}=L(\boldsymbol{\xi}_{t},\boldsymbol{\pi}_{t},t)
\end{eqnarray}
\begin{eqnarray}
\label{}
\zeta_{\ti}=0
\end{eqnarray}
\end{subequations}
and looking for stationary point of the action functional
\begin{eqnarray}
\tilde{A}=\zeta_{\tf}+\int_{\ti}^{\tf}\mathrm{d}t\,
\Bigg{(}
\boldsymbol{\pi}_{t}\cdot\boldsymbol{\dot{\xi}}_{t}+\phi_{t}\dot{\zeta}_{t}
-\bigg{(}\boldsymbol{\pi}_{t}\cdot\boldsymbol{b}(\boldsymbol{\xi}_{t},\boldsymbol{\alpha}_{t},t)
+\phi_{t}L(\boldsymbol{\xi}_{t},\boldsymbol{\alpha}_{t},t)\bigg{)}
\Bigg{)}
\nonumber
\end{eqnarray}
Let us make the simplifying assumption that any pair of trajectory and control variables 
satisfying the boundary have a non-empty open neighborhood where linear variations are well defined.  
Looking for a stationary point of (\ref{Pontryagin:action}) entails considering variations of $\zeta_{t}$
under the constraints $\zeta^{\prime}_{\ti}=\zeta^{\prime}_{\tf}=0$. Then it follows immediately that the stationary
value of the Lagrange multiplier $\phi_{t}$ must satisfy
\begin{eqnarray}
\dot{\bar{\phi}}_{t}=0
\nonumber
\end{eqnarray}
This observation clarifies why the maximum principle is stated for some constant $p_{o}\leq 0$ such that  
$\phi_{t}=p_{o}$. In particular, if $p_{o}\,<\,0 $ we can always rescale it to $p_{o}=-1$ and recover familiar 
form of the Hamilton equations. Moreover, the Maximum principle coincides with the Hamilton form of the stationary
action principle if $\boldsymbol{b}=\boldsymbol{\alpha}_{t}$ and $L$ is quadratic in $\boldsymbol{\alpha}_{t}$.
If instead there exist stationary solutions for $p_{0}=0$, they describe \emph{abnormal controls}.

Abnormal control do not occur in the optimization problem considered in the main text. 
In the push regions where the acceleration is non-vanishing abnormal control drive the Lagrange 
multiplier $\theta_{t}$ away from zero. Thus, they are not compatible with 
the occurrence of switching times between push and no-action regions.
Looking for abnormal control in the no-action region yields the requirement that all Lagrange multipliers vanish
against the hypothesis of the maximum principle. 

\addcontentsline{toc}{section}{Bibliography}

\bibliography{/home/paolo/RESEARCH/BIBTEX/jabref}{} 
\bibliographystyle{myhabbrv} 

\end{document}